\documentclass[twocolumn]{article}

\usepackage{lineno}
\usepackage{graphicx}
\usepackage{dcolumn}
\usepackage{bm}
\usepackage{amsfonts}
\usepackage{amssymb}
\usepackage{amsmath}
\usepackage{bbold}
\usepackage{fix-cm}
\usepackage{mathptmx} 
\usepackage[T1]{fontenc}
\usepackage[colorlinks,allcolors=blue]{hyperref}
\usepackage{color}
\usepackage{caption}
\usepackage[superscript,biblabel]{cite}
\usepackage{authblk}

\usepackage[left=2cm, right=2cm, top=2cm,bottom=2.5cm]{geometry}
\let\OLDthebibliography\thebibliography
\renewcommand\thebibliography[1]{
  \OLDthebibliography{#1}
  \setlength{\parskip}{0pt}
  \setlength{\itemsep}{0pt plus 0.3ex}
}

\begin{document}  

\title{\vspace{-1.cm}\large \bf Impact of pear-shaped fission fragments on mass-asymmetric fission in actinides}  
  

\author[1]{\normalsize Guillaume Scamps\thanks{scamps@nucl.ph.tsukuba.ac.jp}}
\author[2]{C\'edric Simenel\thanks{cedric.simenel@anu.edu.au}}

\affil[1]{Center for Computational Sciences,  University of Tsukuba, Tsukuba 305-8571, Japan}
\affil[2]{Department of Nuclear Physics and Department of Theoretical Physics, Research School of Physics and Engineering\\
 Australian National University, Canberra, Australian Capital Territory 2601, Australia}
\date{\vspace{-0.7cm} \small \today}
     
\maketitle  

{\bf Nuclear fission of heavy (actinide) nuclei results predominantly in asymmetric mass-splits~\cite{andreyev2018}. 
Without quantum shells, which can give extra binding energy to these mass-asymmetric shapes, 
the nuclei would fission symmetrically. The strongest shell effects are in spherical nuclei, 
so naturally the spherical ``doubly-magic'' $\mathbf{^{132}}$Sn nucleus ($\mathbf{Z=50}$ protons), was expected to play a major role.
However, a systematic study of fission has shown that the heavy fragments are distributed around $\mathbf{Z=52}$ to 56~\cite{schmidt2000}, 
indicating that $\mathbf{^{132}}$Sn is not the only driver. 
Reconciling the strong spherical shell effects at $\mathbf{Z=50}$ with the different $\mathbf{Z}$ values of fission fragments observed in nature 
has been a longstanding puzzle~\cite{Schmidt2018}. Here, we show that the final mass asymmetry of the fragments is also determined 
by the extra stability of octupole (pear-shaped) deformations which have been recently confirmed experimentally around $^{144}$Ba ($\mathbf{Z=56}$)~\cite{bucher2016,bucher2017}, 
one of very few nuclei with shell-stabilized octupole deformation~\cite{gaffney2013}. 
Using a modern quantum many-body model of superfluid fission dynamics~\cite{scamps2015a}, 
we found that heavy fission fragments are produced predominantly with $\mathbf{52-56}$ protons, 
associated with significant octupole deformation acquired on the way to fission. 
These octupole shapes favouring asymmetric fission are induced by deformed shells at $\mathbf{Z=52}$ and 56. 
In contrast, spherical ``magic'' nuclei are very resistant to octupole deformation, which hinders their production as fission fragments. 
These findings may explain surprising observations of asymmetric fission of lighter than lead nuclei~\cite{andreyev2010}. 
}

Atomic nuclei are usually found in a minimum of energy ``ground-state'' which may be deformed due to quantum correlations.
Elongation beyond the ground-state costs potential energy until a maximum is reached at the fission barrier. 
Increasing the elongation beyond the fission barrier decreases the potential energy 
and the system follows a fission valley in the ``potential energy surface''  until it breaks into two fragments (scission). 
In the absence of quantum shell effects, all heavy nuclei preferentially fission into two fragments of similar mass (mass-symmetric fission).
However, quantum shells in the fissioning nucleus can result in several valleys to scission. 
These may be mass-symmetric or mass-asymmetric.

\begin{figure}[!h]
\begin{center}
\centering\includegraphics[width=7.7cm]{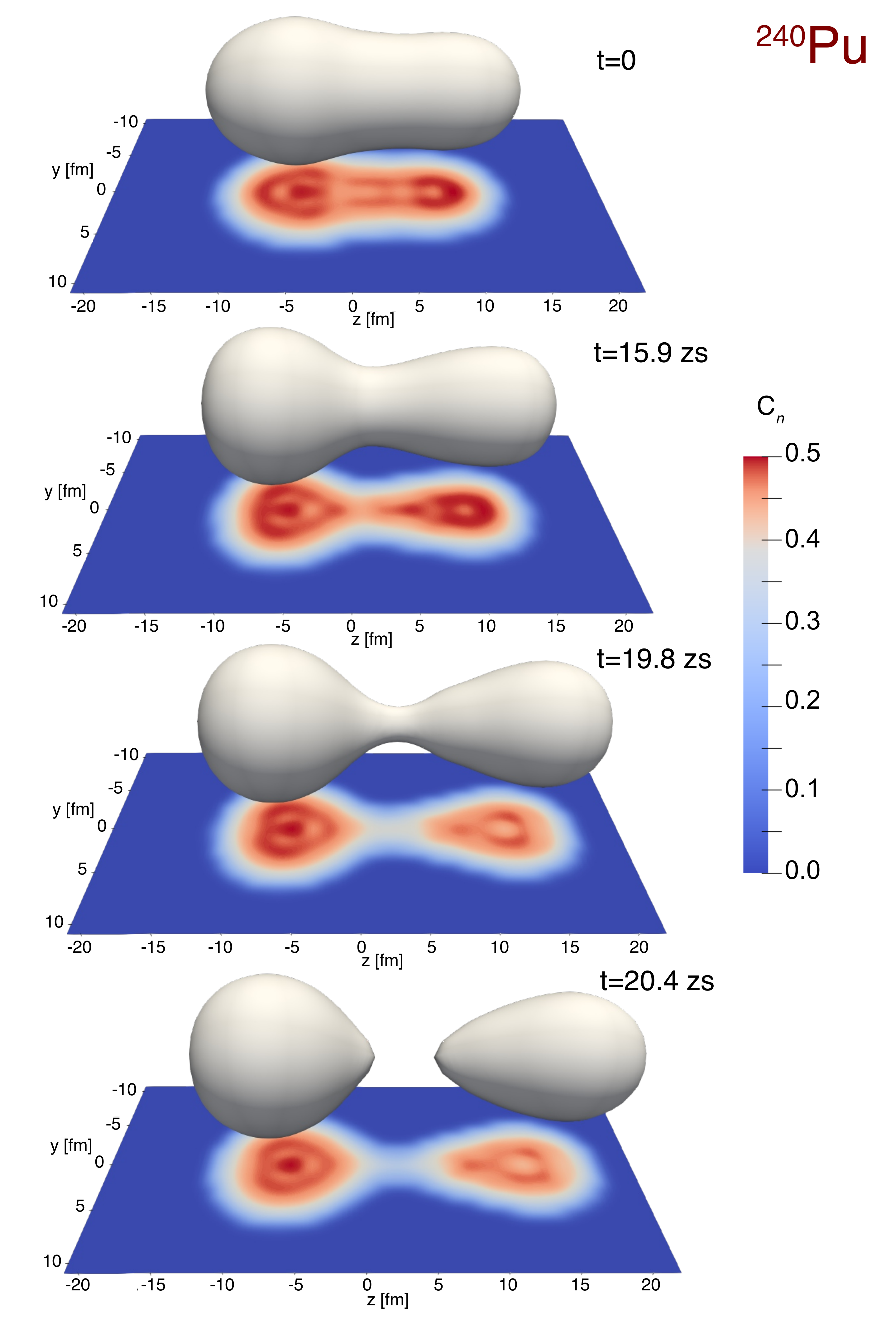}
\end{center}
\caption{{\bf Microscopic calculations of $^{240}$Pu asymmetric fission.}\\
Isodensity surfaces at $0.08$~fm$^{-3}$ (half the saturation density) have been computed from the full microscopic evolution 
and are shown at  different times (see Supplementary Information video). 
The localisation function $C_n$ of the neutrons (see Methods) is shown in the projections. 
Here, scission occurs at $t\simeq20$~zs, i.e., between 3rd and 4th panels.
The quadrupole and octupole deformation parameters (see Methods) at scission are $\beta_2\simeq0.16$, $\beta_3\simeq0.22$ for the heavy fragment (left) and $\beta_2\simeq0.64$, $\beta_3\simeq0.4$ for the light fragment (right).
} 
\label{fig:3d_view}
\end{figure}

\begin{figure*}[!h]
\begin{center}
\includegraphics[width=  8 cm ]{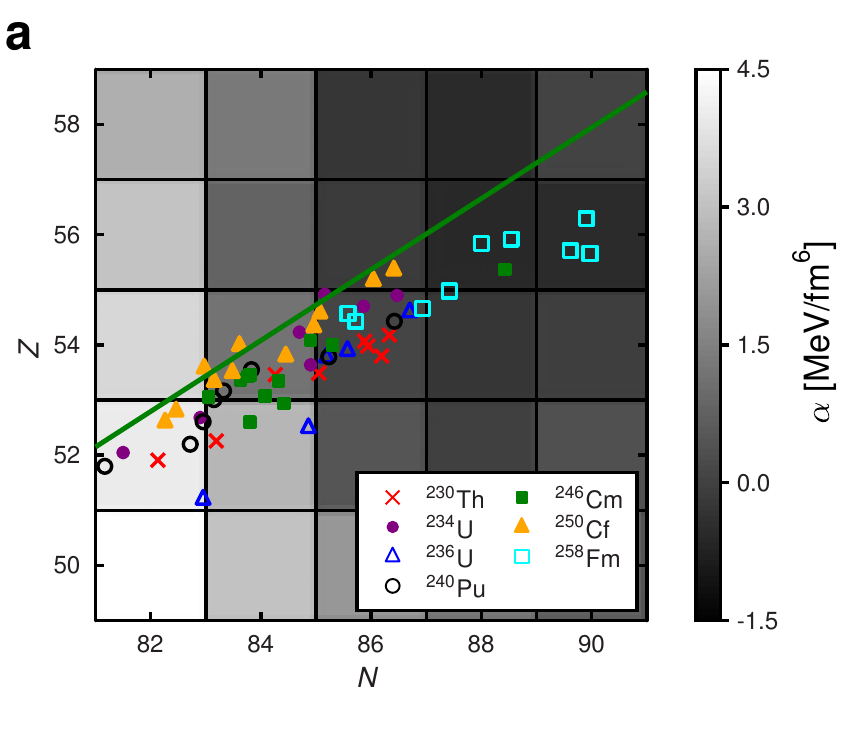}
\hspace{0.5cm}
\includegraphics[width= 7.5cm]{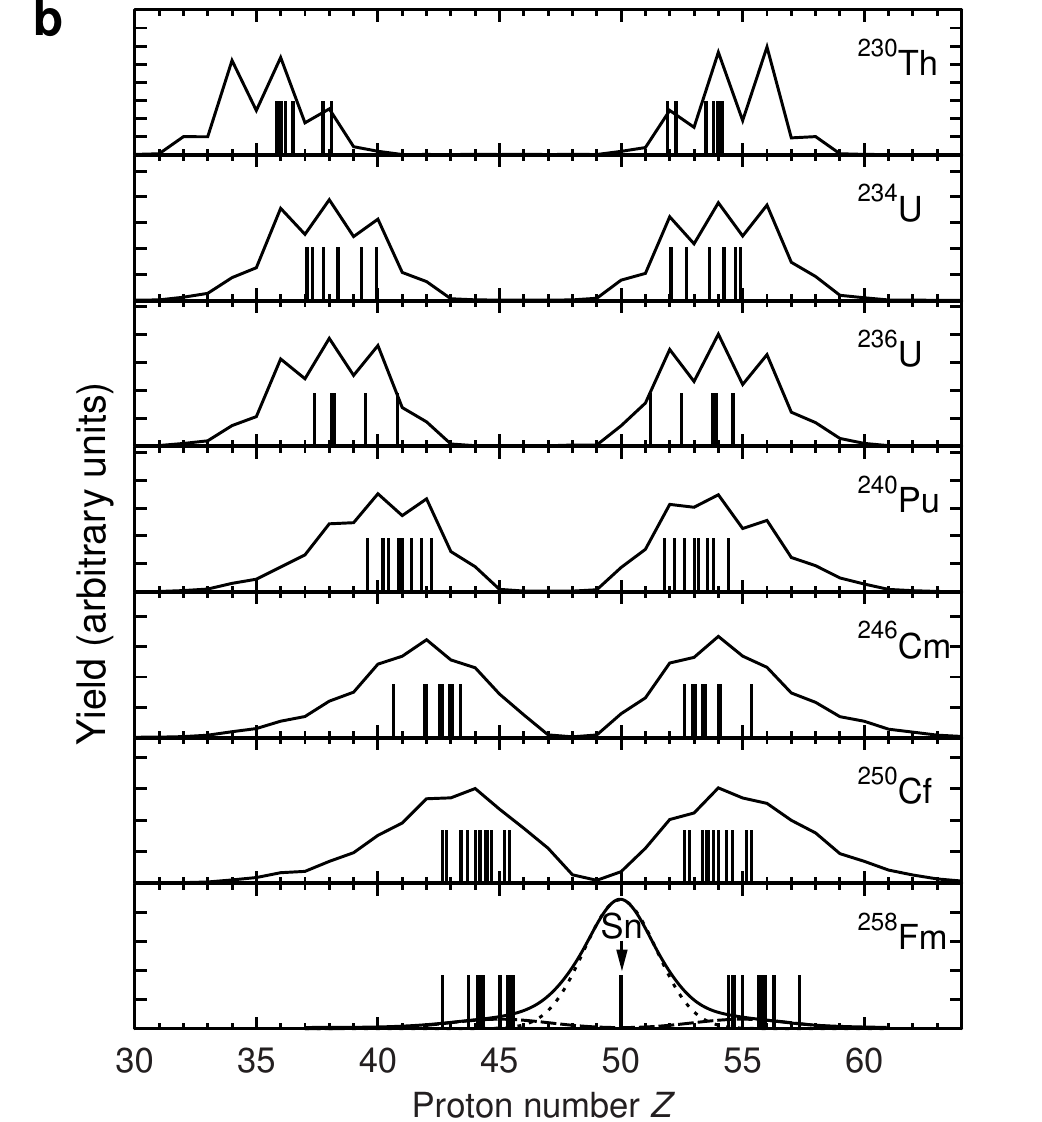}
\end{center}
\caption{{\bf Proton and neutron number distributions in fission fragments.} \\
{\bf a}, Expectation values of the number of protons and neutrons in the heavy fragments for various asymmetric fissions.
The solid line shows the expected positions of fragments with the $N/Z$ neutron-to-proton number ratio values of $^{240}$Pu.
The background grey scale quantifies the resistance to octupole (pear-shaped) deformations in the nuclei as predicted by the Constrained Hartree-Fock+BCS (CHF+BCS) calculations. 
It is obtained from the curvature of the octupole deformation energy 
$\alpha=\lim_{Q_{30}\to0}\frac{E}{Q_{30}^2}$ of the nuclei, where $Q_{30}$ is the octupole moment (see Methods), near their energy minimum at $Q_{30}=0$ (corresponding to the curvature at origin in Fig.~\ref{fig:Q2_Q3}b).
Negative values indicate nuclei likely to exhibit octupole deformations in their ground-state.  
{\bf b}, Our microscopic predictions of expectation values of the  number of protons $\langle Z\rangle$ in the fission fragments (vertical lines) 
are compared with fragment proton number distributions (solid lines) extracted from thermal neutron induced fission  (top 6 panels)  \cite{BROWN20181} 
and $^{258}$Fm spontaneous fission (bottom panel) \cite{hulet1986} experimental results.
Dashed and dotted lines in the bottom panel are Gaussian fits of the asymmetric and symmetric components, respectively. } 
\label{fig:distributions}
\end{figure*}

Although recent progress has been made in describing fission fragment mass distributions with stochastic based approaches \cite{moller2015,sadhukhan2016}, theoretical description of the first stage of fission, from the ground-state deformation to the fission barrier, remains a challenge \cite{schunck2016}.
However, the study of the dynamics along the fission valleys  is now possible 
with the time-dependent energy-density functional approach \cite{simenel2014a,goddard2015} 
including nuclear superfluidity \cite{scamps2015a,bulgac2016,tanimura2017}.
This approach is quantum and fully microscopic in that it follows in time the evolution 
of all single-particle wave-functions (196 for neutrons and 126 for protons in the present calculations) 
which are fully or partly occupied.

Figure~\ref{fig:3d_view} shows the dynamical evolution of isodensity surfaces of $^{240}$Pu 
starting from a configuration in the asymmetric fission valley.
The final state corresponds to a quantum superposition of different repartitions of the number of nucleons between the fragments, 
with  $\langle Z\rangle\simeq53.8$ protons and $\langle N \rangle \simeq85.2$ neutrons in average in the heavy  (left) fragment.
Similar calculations have been performed for $^{230}$Th, $^{234,236}$U, $^{246}$Cm, $^{250}$Cf and $^{258}$Fm actinides 
(Extended Data Tab.~\ref{Tab:all_data} and Extended Data Fig.~\ref{fig:3d_view_scission}).
For each system, a range of initial configurations in the asymmetric fission valley has been considered in order to investigate 
 the influence of the initial elongation (and consequently the initial potential energy) on the final  properties of the fragments.
In the example of Fig.~\ref{fig:3d_view}, scission occurs after $\sim20$~zs (1~zs$=10^{-21}$~s), 
but can reach up to 90~zs depending on the initial configuration. 
Despite large fluctuations in the time to scission, almost all heavy fragments are formed with $Z=52-56$ protons 
(see Fig.~\ref{fig:distributions}), in excellent agreement with experimental data. 
The fact that the number of protons shows little (if any) dependence on the initial elongation
is due to the slow viscous motion of fissioning nuclei which essentially follow the bottom of their fission valley\cite{bulgac2018fission}, 
except near scission where the evolution is expected to be faster \cite{simenel2014a}.

We also observe in Fig.~\ref{fig:3d_view}  and Extended data Fig.~\ref{fig:3d_view_scission} that the fragments are formed with a strong deformation at scission.
This deformation results from the competition between the long-range Coulomb interaction which repels the fragments,
the short range nuclear attraction in the neck between the fragments, and the deformation energy of the fragments.
The latter quantifies the energy cost to deform a fragment, 
which can be particularly large for spherical doubly-magic nuclei like $^{132}$Sn, 
or small for non-magic nuclei which are often deformed in their ground-state.

The strong nuclear force attraction between the fragments is responsible for the neck (see middle panels in Fig.~\ref{fig:3d_view}),
inducing quadrupole (cigar-shaped) and octupole (pear-shaped) deformations of the fragments.
Although the quadrupole deformation is often taken into account in modelling fission \cite{moller2001}, 
the octupole deformation is also important in describing scission configurations properly \cite{carjan2017}.
The neutron localisation function \cite{becke1990} (see Methods), shown as projections in Fig.~\ref{fig:3d_view}, also exhibits strong octupole shapes.  
In addition, the localisation function  is often used to characterise shell structures in quantum many-body systems such as nuclei 
\cite{reinhard2011,jerabek2018} and atoms \cite{jerabek2018}.
For instance, we see in the second panel of Fig.~\ref{fig:3d_view} that the signature shell structures of particular fragments, as well as their deformation, 
are already present $\sim4$~zs before scission. Examples of identification of the pre-fragment as octupole deformed $^{144}$Ba  ($Z=56$)
are given in Extended Data Figs.~\ref{fig:prefrag} and ~\ref{fig:prefrag_syst}.

Our calculations show that, indeed, Sn fragments produced in $^{258}$Fm symmetric fission have  octupole moments 
2 to 3 times smaller at scission than those with $Z\approx55$.
Fragments produced with $Z\approx52$ also exhibit a significant octupole deformation, 
though not as strong as $Z\approx55$ fragments (see Extended data Fig. \ref{fig:Q3_fct_t_3D}).
The case of the doubly magic $^{132}$Sn spherical nucleus is particularly interesting.
On the one side, its production as a fission fragment is usually expected to be favoured 
because of its extra-binding energy originating from spherical shell effects.
On the other side, its deformation (which is inevitable in fission) costs more energy, thus hindering its production as a fission fragment.

A similar interplay between intrinsic deformation of the fragments and their relative motion is well known in  heavy-ion fusion~\cite{dasgupta1998}.
The capture mechanism can naively be seen as the ``reverse'' process to scission occurring at the late stage of nuclear fission.
Therefore, it is not surprising that similar couplings to octupole shapes impact the fission dynamics, 
favouring the formation of fission fragments which exhibit octupole correlations.  
Of course, a similar role is played by the quadrupole couplings 
(see for instance the strong quadrupole deformation of the light fragment in Fig.~\ref{fig:3d_view}).
However, as the majority of nuclei exhibit a quadrupole deformation in their ground-state, 
this cannot be the only reason for the specific number of protons ($Z\sim52-56$) of the heavy fragment in actinide asymmetric fission.

\begin{figure}[!h]
\begin{center}
\includegraphics[width=  \linewidth]{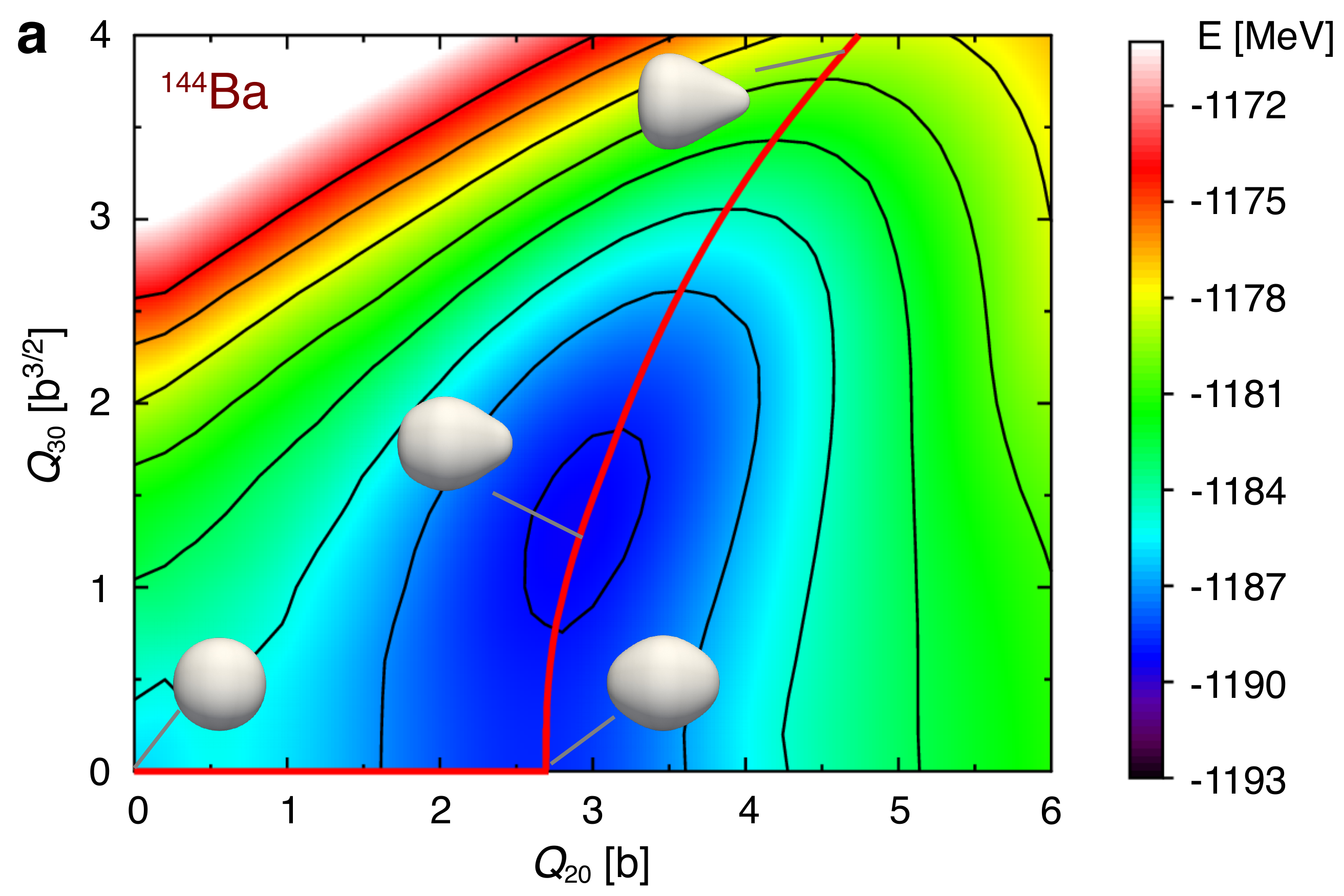}
\includegraphics[width=  \linewidth]{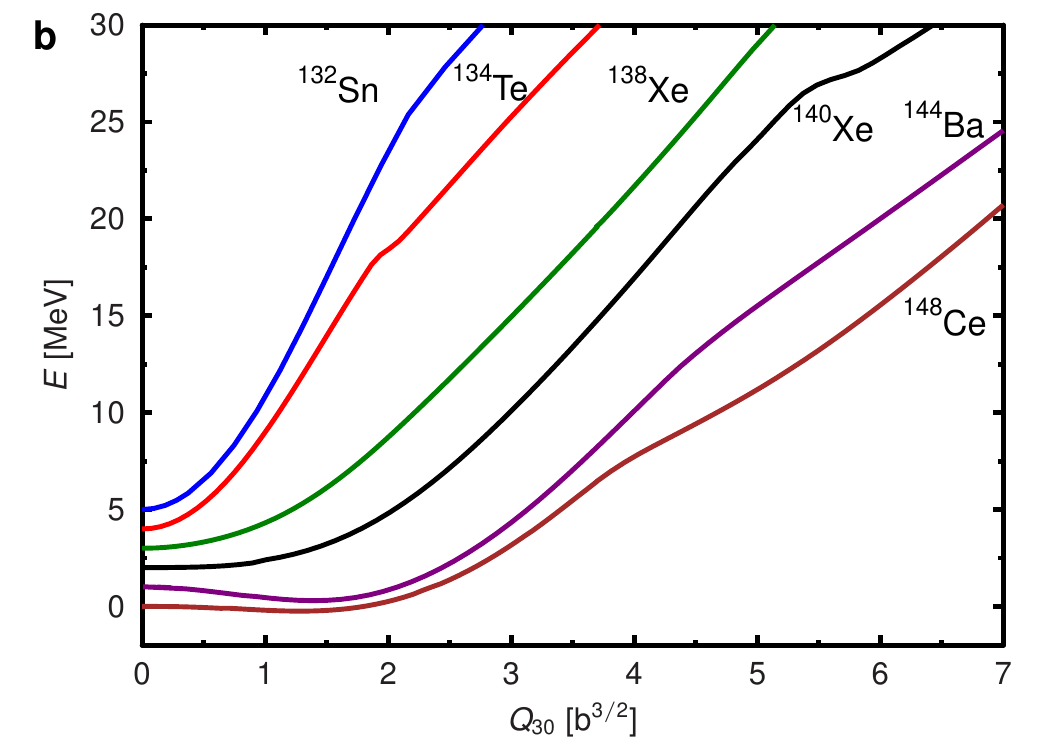}
\end{center}
\caption{{\bf Deformation energy in fission fragments.}\\
{\bf a}, Potential energy surface of $^{144}$Ba. The binding energy using the Sly4d functional is shown as a function of the quadrupole ($Q_{20}$) and octupole ($Q_{30}$) moments (see Methods). 
Binding energy increases from blue to red, with the iso-energy contour lines  separated by an energy of 2~MeV. 
The red thick line represents the minimum energy at a given quadrupole deformation.
{\bf b}, Microscopic calculations of octupole deformation energy (defined as the binding energy minus the binding energy for a spherical nucleus) for several isotopes produced in fission of actinides.
All other multipole moments, including the quadrupole moment, are not constrained. 
Thus, the curves show the minimum of energy for a given octupole moment. 
The reference energy is shifted up by 1~MeV steps for each curve to improve clarity of the figure.
While $^{144}$Ba is predicted with an octupole minimum, 
these octupole correlations disappear for Sn ($Z=50$) magic nuclei. 
} 
\label{fig:Q2_Q3}
\end{figure}

In contrast, fewer nuclei are expected to exhibit octupole deformation in their ground state \cite{butler1996,robledo2011,gaffney2013,butler2016,bucher2016}. 
Recent experiments have confirmed non-ambiguously that this is the case of $^{144}$Ba ($Z=56$) \cite{bucher2016}, 
one possible heavy fragment in  asymmetric fission of actinides.
Nuclei close to  $^{144}$Ba  in the nuclear chart should also exhibit particularly strong octupole correlations
thereby providing a possible explanation to the favoured production of these nuclei in fission. 
The microscopic approach used here automatically incorporates the possibility 
for the nuclei to acquire octupole deformations induced by their underlying quantum shell structure. 
This is illustrated in Fig.~\ref{fig:Q2_Q3}a showing the potential energy surface of $^{144}$Ba 
and predicting the ground-state to be quadrupole and octupole deformed.
According to the deformation energies plotted in Fig.~\ref{fig:Q2_Q3}b, these octupole correlations are present in nuclei with proton and neutron numbers close to $^{144}$Ba and $^{140}$Xe. 
However, they disappear in nuclei close to $^{132}$Sn which, as expected, are found to be resistant against octupole deformations (see also Extended Data Fig. \ref{fig:deform_Q}).

\begin{figure}[!h]
\begin{center}
\includegraphics[width=  \linewidth]{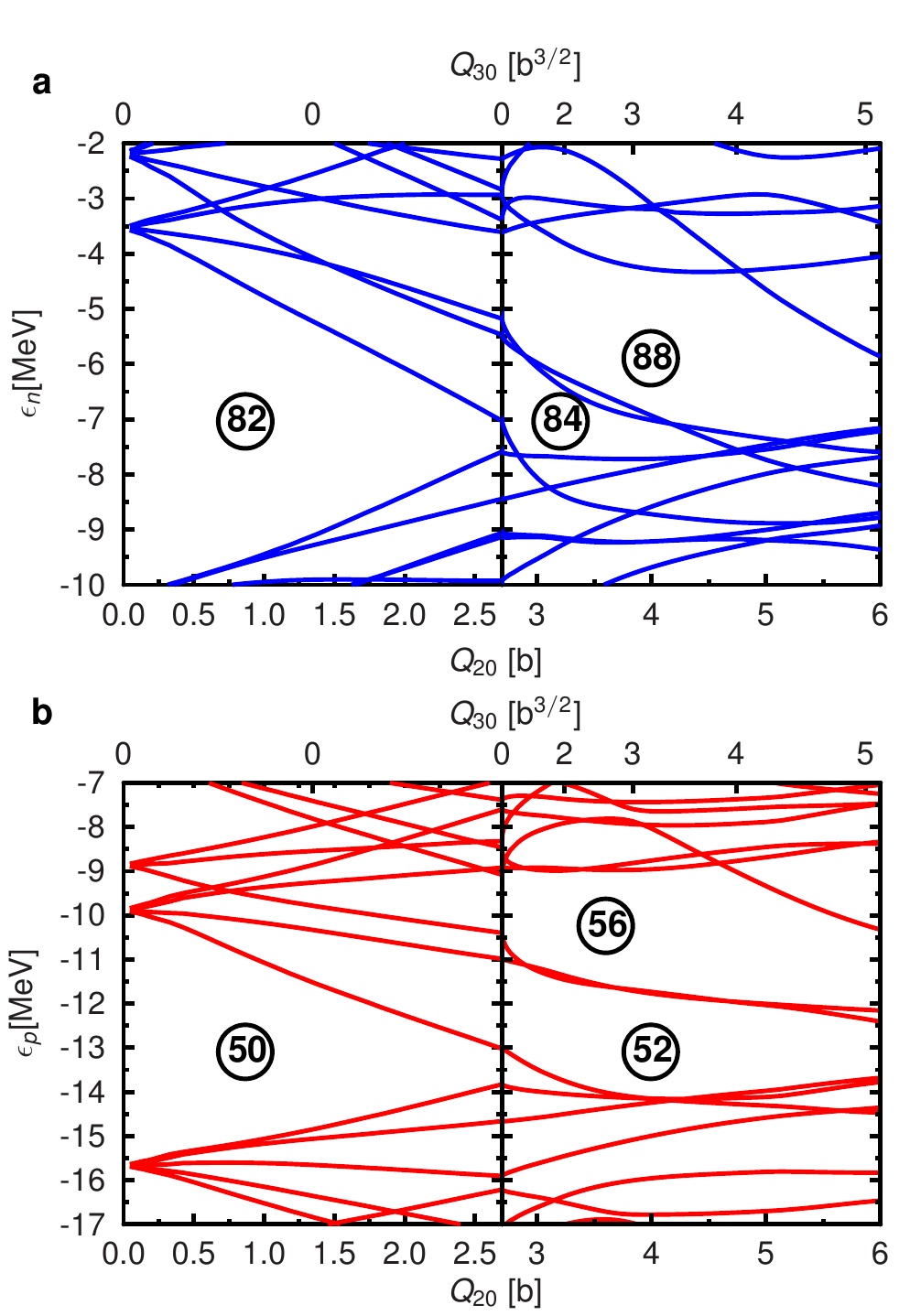}
\end{center}
\caption{ {\bf  Evolution of single-particle energies with deformation. }\\
 Neutron ({\bf a}) and proton ({\bf b}) single-particle energies as a function of the quadrupole (lower scale) 
and octupole (upper scale)  moments in $^{144}$Ba following the path of the red solid line in Fig. \ref{fig:Q2_Q3}a. 
The numbers in the energy gaps correspond to the number of particles which can possibly occupy the single-particle states below the energy gap.
50 and 82 are spherical ``magic'' numbers while 52, 56, 84, and 88 are associated with deformed energy gaps. See also\cite{leander1985}.} 
\label{fig:struct}
\end{figure}

The origin of the octupole correlations can be understood from the shell structure of the nuclei. 
Like in atoms, the properties of the spectrum of single-particle energy levels are often found to be responsible 
for quantum shell effects in atomic nuclei. 
For instance, spherical nuclei with a ``magic'' number of protons and neutrons (the analog of noble gas) have all their proton and neutron levels filled 
below large energy gaps of few MeV, above which the levels are empty. 
As a result, the nucleus is difficult to excite and acquires an extra binding energy from these quantum shell effects. 
However, the single-particle energy spectrum changes with the deformation of the nucleus,
in such a way that spherical energy gaps disappear while other gaps may appear in deformed nuclei, stabilising  their shape.
The calculated neutron and proton single-particle energies in $^{144}$Ba are shown in Fig.~\ref{fig:struct} 
as a function of quadrupole and octupole deformation, 
following the path of minimum energy (solid red line in Fig.~\ref{fig:Q2_Q3}a). 
The large energy gaps at the spherical point ($Q_{20}=Q_{30}=0$) correspond to the ``magic'' numbers $Z=50$ and $N=82$
which are responsible for the shell effects in $^{132}$Sn. 
However, we observe a closure of these spherical energy gaps and the opening of $Z=52,56$ and $N=84,88$ deformed shell gaps 
which survive for a large range of octupole deformations. 
The matching between the positions of the proton deformed gaps $Z=52,56$ in Fig.~\ref{fig:struct}b and  the proton numbers in the heavy fragments 
(see Fig.~\ref{fig:distributions}b) is striking.
The fact that the final mass asymmetry can be explained by octupole deformed shells without invoking the $Z=50$ spherical gap is surprising. Note that the observed importance of octupole correlations at scission does not exclude a possible contribution of magic shells in the early stage of fission, e.g., via the creation of valleys in the potential energy surface.
Several competing effects are then expected to be at play in the formation of these valleys. 
These include the spherical and octupole shell gaps of the heavy fragment, in addition to other possible deformed shell gaps in the light one.
In fact, amongst all actinides studied experimentally, only $^{258}$Fm is clearly dominated by  $Z=50$ spherical shell effects \cite{hulet1986}, 
producing two symmetric Sn fragments responsible for the narrow peak 
in its fission fragment charge distribution (see of Fig.~\ref{fig:distributions}b), 
although it also exhibits  a weaker asymmetric mode with a heavy fragment around $Z\approx55$.
However, asymmetric fission  dominates in $^{256}$Fm and lighter fermium isotopes  \cite{unik1974}, 
confirming the weak influence of $Z=50$ in the formation of fission fragments. 

The total kinetic energy (TKE) of the final fragments is another important observable which can be used to characterise fission properties.
It is indeed related to the elongation of the system at scission, and thus to the deformation of the fragments. 
For instance, a large elongation at scission reduces the Coulomb repulsion between the fragments, 
therefore leading to a smaller TKE when the fragments are far away and the Coulomb energy has been converted into kinetic energy.  
In contrast, a compact system with near spherical fragments at scission 
is usually associated to larger TKE. 
The TKE  computed from our microscopic simulations, following the method introduced in Ref.~\cite{simenel2014a}, 
are in very good agreement with experimental data \cite{bockstiegel2008,caamano2015} (see Extended Data Tab.~\ref{Tab:all_data} and Fig. \ref{fig:TKE_fct_Z}).
In particular, most of our calculated TKE are found to be smaller for  fission leading to  $Z\simeq55$ than $Z\simeq52$ fragments, 
indicating that the latter is more compact at scission.  
The smaller TKE for $Z\simeq55$ is interpreted as an effect of the larger deformation in the heavy fragments.

The fact that both the mass-asymmetric fission of actinides and the TKE measured experimentally can be explained 
by our time-dependent microscopic calculations gives us confidence in the fission dynamics predicted by these calculations, 
and, in particular, in the major role played by octupole deformation of the heavy fragments. 
The octupole deformation in $Z_{\rm light}\simeq34$ and $N_{\rm heavy}=56$ nuclei \cite{butler2016} could also explain mass-asymmetric fission found experimentally \cite{andreyev2010} in lighter systems.
Other properties of fission will also be investigated in the future, such as the excitation energy of the fragments 
and the number of neutrons emitted during the fission process.

\noindent
{\bf Supplementary Information} \\
A video showing calculations of $^{240}$Pu fission dynamics is available in the online version of the paper.\\

\noindent
{\bf Acknowledgements}\\
B. Jurado, A. Chatillon and F. Farget are thanked for useful discussions at the early stage of this work.
We are grateful to D. J. Hinde for his continuous support to this project.
We thank M. Caama\~no for providing references to experimental data.
B. Jurado and D. J. Hinde are also thanked for their careful reading of the manuscript.
This work has been supported by the Australian Research Council under Grant No. DP160101254. 
The calculations have been performed in part at the NCI National Facility in Canberra, Australia, 
which is supported by the Australian Commonwealth Government, 
 in part using the COMA system at the CCS in University of Tsukuba supported by the HPCI Systems Research Projects (Project ID hp180041), and using the Oakforest-PACS at the JCAHPC in Tokyo supported in part by Multidisciplinary Cooperative Research Program in CCS, University of Tsukuba.\\

\noindent
{\bf Author Contributions}\\
 G.S. and C.S. conceived the project. G.S. performed the numerical simulations. G.S. and C.S. discussed the results. C.S. wrote the manuscript.\\

\noindent
{\bf Author Information} \\
Correspondence and requests for materials should be addressed to G.S. (scamps@nucl.ph.tsukuba.ac.jp). The authors declare no competing interests.\\

\noindent
{\bf Reprints and permissions information} is available at http://www.nature.com/ reprints.

 
\section*{METHODS}

{\bf Constrained and time-dependent Hartree-Fock calculations with dynamical Bardeen-Cooper-Schrieffer pairing correlations (CHF+BCS and TDBCS)} \\
\noindent
The calculations were done in a three-dimensional
Cartesian geometry with one plane ($y=0$) of symmetry using the code of Ref.~\cite{scamps2015a} and using the
Skyrme SLy4d energy density functional\,\cite{kim1997} with 
surface pairing interaction with interaction strength $V_0^{nn}$ = 1256~MeV$\cdot$fm$^3$  and $V_0^{pp}$ = 1462~MeV$\cdot$fm$^3$ 
in the neutron and proton pairing channels, respectively \cite{scamps2013a}. The effect of the choice of the pairing interaction and functional on the octupole deformation is shown on Extended Data Fig. \ref{fig:effect_interact}.  The cut-off function and parameters are the same as in Ref \cite{scamps2013a}.
The three-dimensional Poisson equation for the Coulomb potential
is solved and the Slater approximation is used for the Coulomb exchange term.
The TDBCS calculation are first performed in a spacial grid of dimension $L_x \times 2L_y \times L_z $ =  19.2 $\times$  19.2 $\times$ 40 fm$^3$ until the fragments reach a relative distance of 17~fm between their centres of mass.  Then the system is described in a larger box with $L_z =$ 56 fm.
The mesh spacing is $0.8$~fm in all directions and the time step between two time iterations is $1.5\times10^{-24}$~s.

 Following the technique of Ref.~\cite{scamps2015a}, the initial configurations for the time-dependent calculations 
 are generated by  constrained Hartree-Fock+BCS (CHF+BCS) calculations  
 with a combination of quadrupole and octupole constraints used to control the elongation and 
 the mass asymmetry, respectively.
 The fission occurs along the $z-$axis. 
 The other multipole moments are not constrained. 
 Once in the asymmetric fission valley, the octupole constraint is released to allow the system to find a local minimum of energy for a given quadrupole deformation~\cite{scamps2015a}. 
A range of 5 to 11 initial quadrupole moments between $Q_{20}\simeq34$ and 72~b (1~barn~$\equiv1$~b~$=10^{-28}$~m$^2$) 
have been considered  for each system with steps of $1-10$~b (see Extended Data Tab.~\ref{Tab:all_data}). 

The CHF+BCS calculations for Figs. \ref{fig:distributions}a (background), \ref{fig:Q2_Q3}, \ref{fig:struct} and Extended Data Fig. \ref{fig:deform_Q} are done with a modified version of the \textsc{ev8} code \cite{bonche2005} where only one plane of symmetry ($y=0$) is used. The spatial  grid for those calculations is of dimension $L_x \times 2L_y \times L_z $ =    22.4 $\times$  22.4 $\times$ 25.6 fm$^3$. 

Pairing correlations are treated dynamically, i.e., without the frozen occupation approximation. Although, the BCS approximation violates the continuity equation \cite{scamps2012}, the spurious transfer of particle after scission turns out to be very small in the case of fission\cite{tanimura2017}.
Compared to the more general Bogoliubov treatment of pairing correlations, the BCS approximation has the advantage to reduce substantially the computational need, while the resulting fission dynamics (saddle to scission times) and properties of the fission fragments (mass, charge and TKE) are very similar with both Bogoliubov\cite{bulgac2016,bulgac2018fission} and BCS dynamical pairing\cite{scamps2015a,tanimura2017}. See also Ref.\cite{simenel2018} for a review.

\noindent
{\bf Multipole moments and deformation parameters}\\
The quadrupole moment is expressed as 
 $Q_{20}=\sqrt{\frac{5}{16\pi}}\int d^3r \,\rho(\mathbf{r}) (2z^2-x^2-y^2)$
 and the octupole moment as 
 $Q_{30}=\sqrt{\frac{7}{16\pi}}\int d^3r \,\rho(\mathbf{r}) [2z^3-3z(x^2+y^2)]$,
 where $\rho(\mathbf{r})$ is the density of nucleons. 
The $\beta_2$ and $\beta_3$ deformation parameters are obtained from the quadrupole and octupole moments following,
\begin{align}
\beta_{\lambda} = \frac{4 \pi }{3 A (r_0 A^{1/3})^{\lambda} }  Q_{\lambda 0},
\end{align}
with $r_0$~=~1.2~fm.

\noindent
{\bf Fermion localisation function} \\
The localisation function is computed as \cite{becke1990}
\begin{align}
	{ C}_{q \sigma}( {\bf r})  = \left[  1+ \left( \frac{\tau_{q \sigma } \rho_{q \sigma }  - \frac14 | \nabla \rho_{q \sigma } |^2  - {\bf j}^2_{q \sigma }  }{   \rho_{q \sigma }  \tau^{TF}_{q \sigma }  } \right)^2 \right]^{-1},
\end{align}
with the nucleon ($\rho_{q \sigma }$), kinetic ($\tau_{q \sigma }$) and current (${\bf j}_{q \sigma }$) densities defined as
\begin{align}
	\rho_{q \sigma } ({\bf r}) &= \sum_{\alpha \in q}  n_{\alpha}  \varphi^*_{\alpha} ({\bf r} \sigma)\varphi_{\alpha} ({\bf r} \sigma),\\
	\tau_{q \sigma } ({\bf r}) &= \sum_{\alpha \in q}  n_{\alpha} \left| \nabla \varphi_{\alpha} ({\bf r} \sigma) \right|^2, \\
	{\bf j}_{q \sigma } ({\bf r}) &= \sum_{\alpha \in q}  n_{\alpha}   {\rm Im} \left[  \varphi^*_{\alpha} ({\bf r} \sigma) \nabla \varphi_{\alpha} ({\bf r} \sigma) \right], 
\end{align}
where $q$ stands for neutron or proton and $\sigma$ is the spin.
$\tau^{TF}$ is the Thomas-Fermi approximation of the kinetic density.
To study the inner core of the nuclei, we suppress the localisation function on the surface of the fragments by applying the transformation\cite{zhang2016},
\begin{align}
	{ C}_{q \sigma}( {\bf r})  \rightarrow { C}_{q \sigma}( {\bf r})  \frac{\rho_{q \sigma } ({\bf r})}{ {\rm max} \left[  \rho_{q \sigma } ({\bf r}) \right]}.
\end{align}
The neutron ($q=n$) localisation function is obtained by averaging over the spin $\sigma$.



\renewcommand{\refname}{ }

\

\noindent
{\bf Data availability}
The data sets generated during the current study are available from the corresponding author on reasonable request.


\section*{EXTENDED DATA}

 \setcounter{figure}{0} 
 
 \renewcommand{\figurename}{Extended Data Figure}
 \renewcommand{\tablename}{Extended Data Table}
 
 \begin{table*}[h]
\caption{ {\bf Results of the Time-Dependent Hartree-Fock calculations with Bardeen-Cooper Schrieffer dynamical pairing correlations (TDBCS).}\\ 
Fissioning nucleus, simulation number,  quadrupole moment $Q_{20}$ and potential energy $E_0=E_{\rm ini}-E_{\rm g.s.}$ of the fissioning system in the initial condition of the TDBCS calculation, 
time $T$ to reach scission,  average proton and neutron numbers $\langle Z_H\rangle$ and $\langle N_H\rangle$ in the heavy fragment, and  total kinetic energy (TKE) in MeV of the fragments.}
\centering
\setlength{\tabcolsep}{0.28em}
\begin{tabular}{cccccccc}
\hline
Nucl. &\# & Q$_{20}$ [b] & E$_0$[MeV] &   T [zs] &   $\langle Z_H\rangle$    &   $\langle N_H\rangle$     & TKE  \\
 \hline
$^{230}$Th & 1 & 34.7 & 4.04 & 50.4 & 53.46 & 84.26 & 159.2 \\
 & 2 & 37.8 & 3.16 & 55.3 & 51.91 & 82.13 & 170.0 \\
 & 3 & 41.0 & 1.78 & 23.0 & 53.49 & 85.05 & 157.0 \\
 & 4 & 50.5 & -0.53 & 13.2 & 54.18 & 86.33 & 154.7 \\
 & 5 & 53.6 & -1.08 & 10.4 & 52.26 & 83.19 & 164.8 \\
 & 6 & 56.8 & -1.72 & 7.9 & 53.8 & 86.19 & 155.2 \\
 & 7 & 59.9 & -2.28 & 6.3 & 53.98 & 85.94 & 154.8 \\
 & 8 & 63.1 & -2.7 & 6.0 & 54.07 & 85.88 & 155.0 \\
 \hline
$^{246}$Cm & 1 & 41.0 & 4.41 & 48.2 & 53.34 & 84.32 & 185.7 \\
 & 2 & 42.6 & 3.13 & 46.9 & 52.6 & 83.8 & 182.8 \\
 & 3 & 50.5 & -0.8 & 23.2 & 55.37 & 88.43 & 169.6 \\
 & 4 & 52.1 & -2.14 & 15.4 & 54.08 & 84.9 & 182.8 \\
 & 5 & 53.6 & -2.93 & 37.4 & 53.36 & 83.63 & 181.2 \\
 & 6 & 56.8 & -4.33 & 53.9 & 53.46 & 83.8 & 185.6 \\
 & 7 & 58.3 & -5.04 & 58.9 & 53.06 & 83.05 & 176.9 \\
 & 8 & 59.9 & -5.24 & 17.4 & 54 & 85.3 & 180.0 \\
 & 9 & 61.5 & -5.91 & 15.1 & 52.94 & 84.42 & 175.2 \\
 & 10 & 63.1 & -6.2 & 6.8 & 53.07 & 84.08 & 183.2 \\
 & 11 & 64.7 & -6.72 & 6.9 & 53.43 & 83.77 & 182.1 \\
 \hline
$^{250}$Cf & 1 & 44.2 & 3.14 & 58.5 & 53.34 & 83.15 & 182.8 \\
 & 2 & 45.8 & 2.03 & 45.5 & 53.5 & 83.48 & 187.6 \\
 & 3 & 47.3 & 1.82 & 29.9 & 52.8 & 82.46 & 207.7 \\
 & 4 & 48.9 & 0.32 & 50.4 & 55.17 & 86.04 & 183.7 \\
 & 5 & 50.5 & -0.66 & 86.9 & 53.58 & 82.97 & 189.1 \\
 & 6 & 52.1 & -2.22 & 90.7 & 53.99 & 83.61 & 194.9 \\
 & 7 & 56.8 & -5.01 & 24.8 & 54.58 & 85.07 & 183.2 \\
 & 8 & 58.4 & -5.8 & 23.1 & 52.6 & 82.26 & 186.8 \\
 & 9 & 59.9 & -6.06 & 25.0 & 55.36 & 86.41 & 180.1 \\
 & 10 & 61.5 & -6.82 & 31.2 & 54.33 & 84.96 & 179.2 \\
 & 11 & 63.1 & -7.34 & 9.2 & 53.8 & 84.45 & 188.2 \\
\hline
\end{tabular}
\begin{tabular}{cccccccc}
\hline
Nucl. &\# & Q$_{20}$ [b] & E$_0$[MeV] &   T [zs] &   $\langle Z_H\rangle$    &   $\langle N_H\rangle$     & TKE  \\
 \hline
$^{234}$U & 1 & 41.0 & 1.94 & 19.9 & 52.05 & 81.5 & 177.2 \\
 & 2 & 44.2 & 0.89 & 30.8 & 54.92 & 85.16 & 164.2 \\
 & 3 & 47.3 & 0.39 & 13.6 & 54.23 & 84.7 & 156.6 \\
 & 4 & 50.5 & -0.39 & 32.4 & 54.7 & 85.86 & 158.5 \\
 & 5 & 56.8 & -2.55 & 28.0 & 52.68 & 82.9 & 173.0 \\
 & 6 & 59.9 & -3.73 & 9.5 & 54.9 & 86.47 & 161.2 \\
 & 7 & 63.1 & -4.72 & 7.6 & 53.64 & 84.9 & 164.1 \\
 \hline
$^{236}$U & 1 & 41.0 & 3.03 & 27.9 & 51.2 & 82.95 & 176.6 \\
 & 2 & 47.3 & 0.28 & 30.6 & 53.9 & 85.57 & 165.0 \\
 & 3 & 50.5 & -0.53 & 32.0 & 53.78 & 85.19 & 163.6 \\
 & 4 & 56.8 & -2.5 & 13.0 & 54.6 & 86.7 & 161.3 \\
 & 5 & 59.9 & -3.5 & 13.2 & 52.5 & 84.86 & 164.5 \\
 \hline
$^{240}$Pu & 1 & 45.4 & 1.46 & 20.1 & 53.79 & 85.23 & 171.5 \\
 & 2 & 46.7 & 0.8 & 16.1 & 53.17 & 83.32 & 181.8 \\
 & 3 & 50.5 & -1.16 & 89.9 & 51.8 & 81.17 & 181.8 \\
 & 4 & 53.0 & -2.13 & 22.5 & 52.61 & 82.95 & 177.9 \\
 & 5 & 56.8 & -3.5 & 16.0 & 53.01 & 83.15 & 177.2 \\
 & 6 & 59.3 & -4.3 & 18.0 & 53.55 & 83.83 & 178.4 \\
 & 7 & 63.1 & -5.31 & 22.1 & 54.43 & 86.42 & 163.6 \\
 & 8 & 71.9 & -7.8 & 3.4 & 52.2 & 82.72 & 179.9 \\
 \hline
$^{258}$Fm & 1 & 46.3 & -1.80 & 49.3 & 54.43 & 85.72 & 190.3 \\
 & 2 & 58.7 & -6.48 & 23.2 & 54.57 & 85.58 & 186.7 \\
 & 3 & 61.9 & -8.53 & 18.3 & 57.35 & 91.86 & 180.3 \\
 & 4 & 64.9 & -10.60 & 14.1 & 55.92 & 88.54 & 182.2 \\
 & 5 & 68.0 & -12.25 & 12.4 & 56.29 & 89.90 & 178.8 \\
 & 6 & 71.2 & -13.59 & 9.2 & 55.84 & 88.00 & 181.2 \\
 & 7 & 74.3 & -14.75 & 10.7 & 54.98 & 87.42 & 183.4 \\
 & 8 & 80.5 & -16.79 & 7.3 & 54.66 & 86.93 & 182.0 \\
 & 9 & 86.7 & -18.60 & 5.7 & 55.66 & 89.97 & 177.8 \\
 & 10 & 89.8 & -19.29 & 5.6 & 55.71 & 89.61 & 176.4 \\
 \hline
\end{tabular}
\label{Tab:all_data}
\end{table*}


\clearpage

 \begin{figure*}[!h]
\begin{center}
\includegraphics[width=8.7cm]{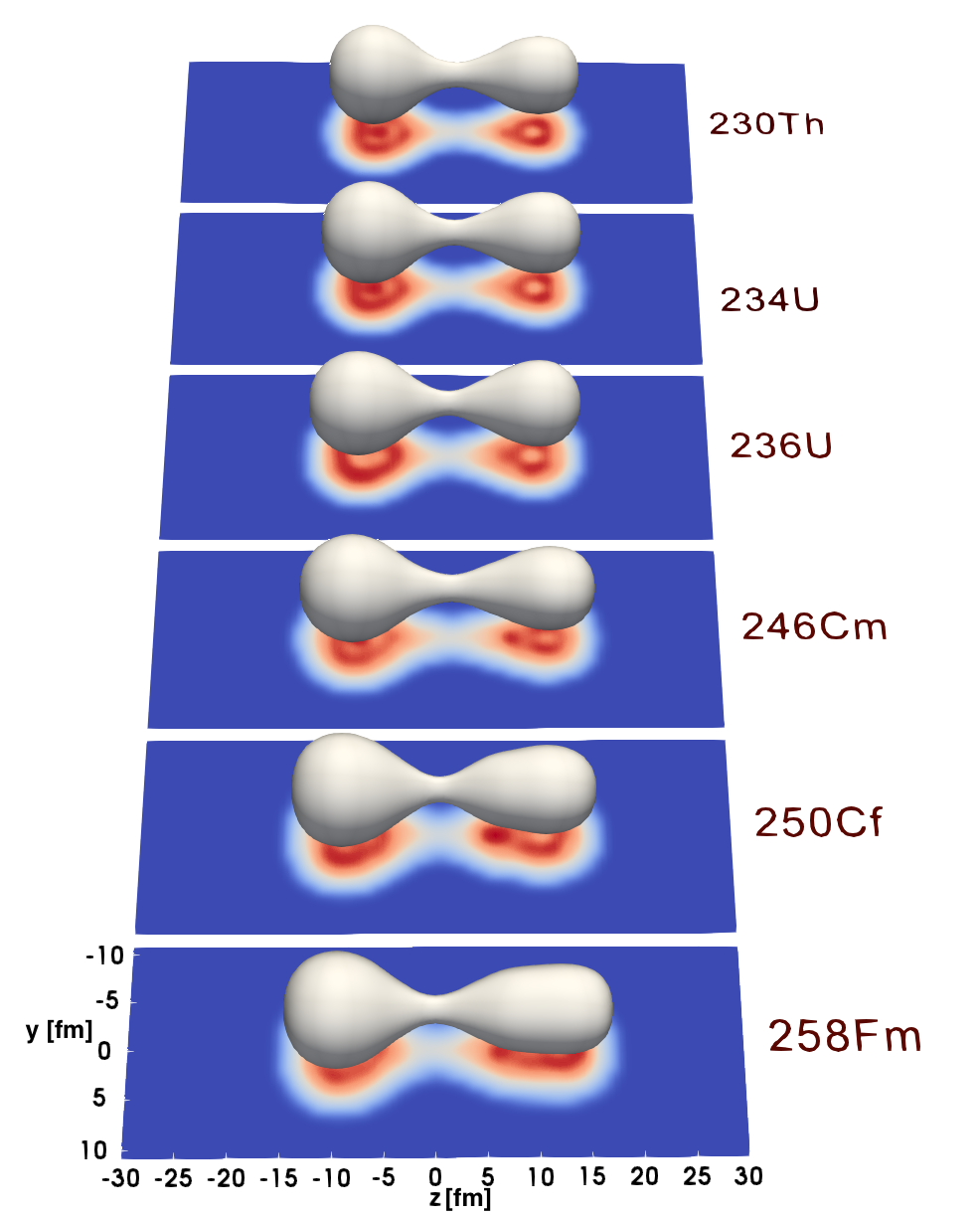}
\end{center}
\caption{{\bf Scission configurations.}\\
 The isodensity surface and the neutron localisation are shown just before scission ($\sim0.1$~zs before the neck breaks)
for calculations with different actinides in their asymmetric fission valley. 
At scission, all the heavy fragments (left) have octupole deformation parameters (see Methods)  $\beta_3\simeq0.25\pm0.02$ and $\beta_2=$0.15-0.27.
These fragments are much more deformed than in the case of $^{258}$Fm symmetric fission (see Extended Data Fig.~\ref{fig:Q3_fct_t_3D}) in which case the symmetric Sn fragments are formed  with $\beta_3\simeq0.11$ at scission. Note that the light fragments have also an octupole deformation with  $\beta_3\simeq0.3-0.4$ and a quadrupole deformation with $\beta_2=0.4-0.8$. Such large quadrupole deformations of the light fragment are often found at scission in microscopic calculations (see, e.g., Fig. 4 of Ref.\cite{sadhukhan2017}).
} 
\label{fig:3d_view_scission}
\end{figure*}

\begin{figure*}[!h]
\begin{center}
\includegraphics[width= 0.6 \linewidth]{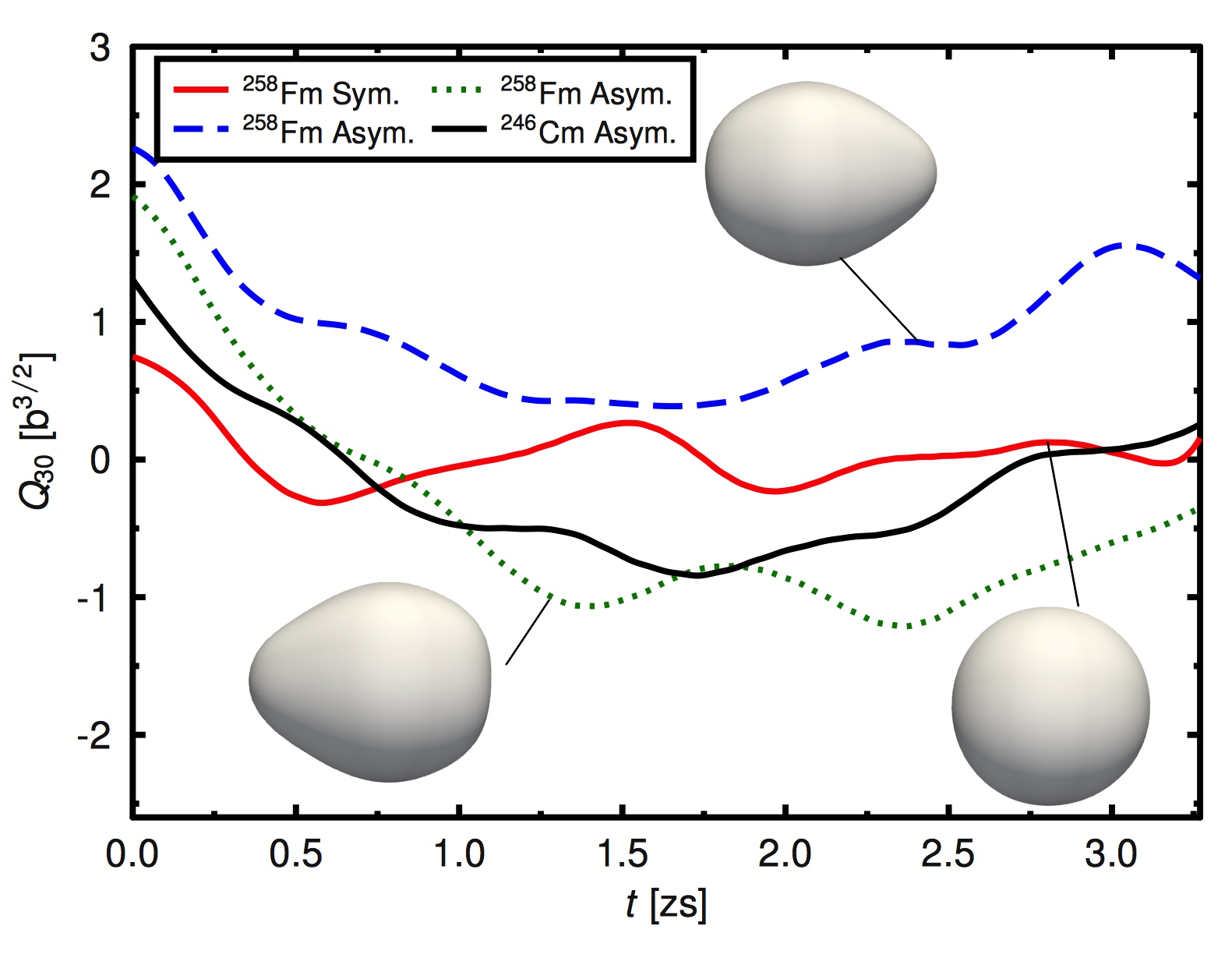}
\end{center}
\caption{ {\bf Octupole deformation after scission.}\\
The octupole moment (see Methods) in the heavy fragment is shown as a function of time, 
with a time reference ($t=0$) corresponding to the time at which scission occurs in the calculations. 
In $^{258}$Fm asymmetric fission, the heavy fragment (with $Z\approx55$) starts with a strong octupole deformation 
(corresponding to deformation parameters $\beta_3\simeq0.25$ at $t=0$) and remains octupole deformed, with possibly different orientations
(blue dashed and green dotted lines).
The fragment with $Z\approx52$ resulting from $^{246}$Cm fission (black solid line) also exhibits a significant, yet smaller, deformation ($\beta_3=0.19$ at $t=0$).
In contrast, $^{258}$Fm symmetric fission produces Sn fragments with a much smaller octupole moment (corresponding to $\beta_3\simeq0.11$ at $t=0$)
which oscillates around $Q_{30}=0$ (red solid line).  
These results are compatible with the octupole deformation energy calculated in Fig.~\ref{fig:Q2_Q3}b which shows that $^{138,140}$Xe ($Z=54$) and $^{144}$Ba ($Z=56$) 
are  less resistant to octupole deformation than $^{134}$Te ($Z=52$) and $^{132}$Sn ($Z=50$).} 
\label{fig:Q3_fct_t_3D}
\end{figure*}

 \begin{figure*}[t]
\includegraphics[width= \linewidth ]{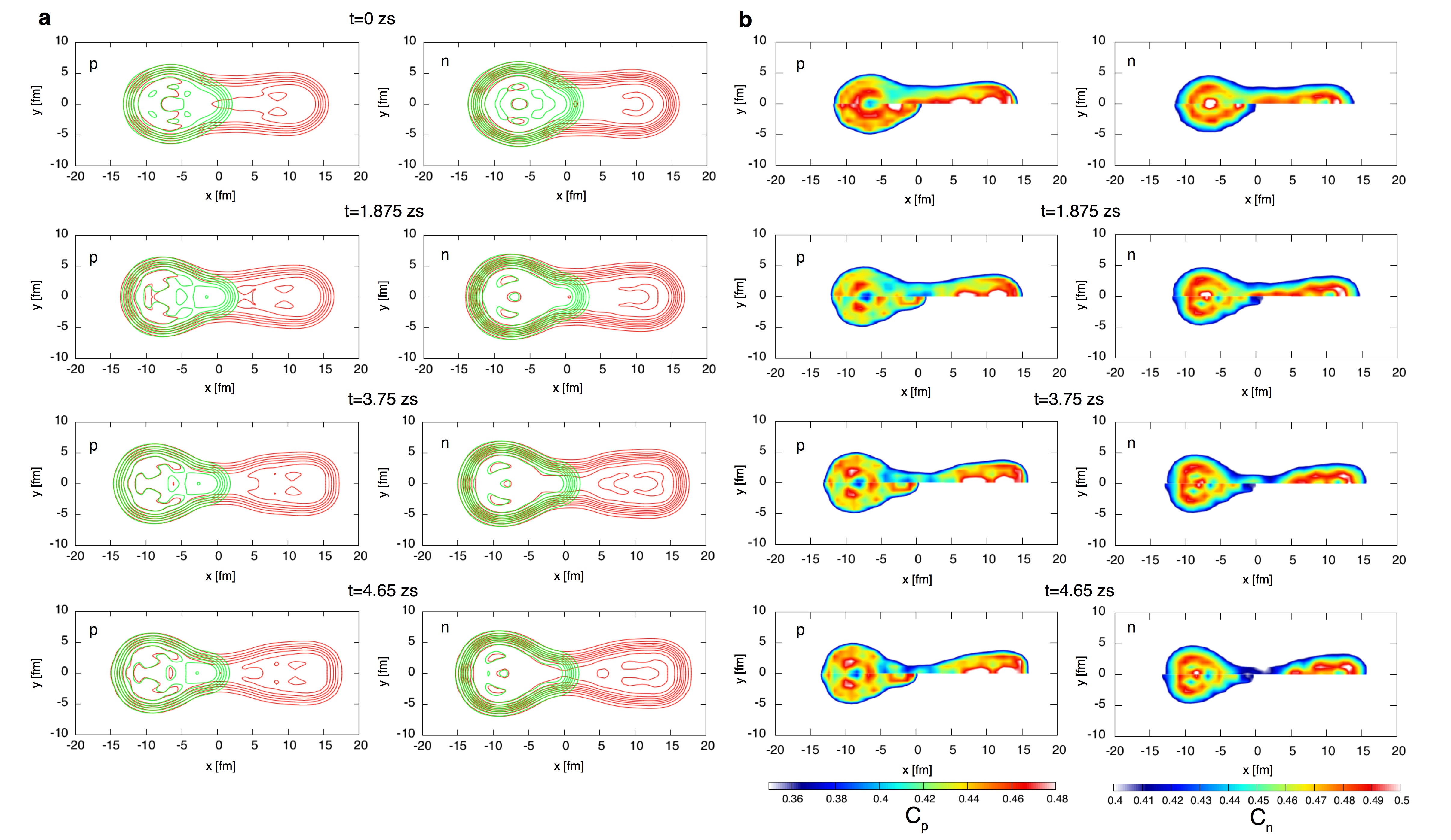}
\caption{{\bf Identification of the heavy pre-fragment in $^{258}$Fm asymmetric fission.  }\\
{\bf a},~The heavy pre-fragment is identified from its density contour using the technique of Ref. \cite{Warda12} without the assumption of a reflection symmetry in the pre-fragment. 
Proton (left column) and neutron (right column) densities are shown with a difference between contour lines of 0.01~fm$^{-3}$.
The fissioning $^{258}$Fm  asymmetric system (red lines, corresponding to calculation \#8 in Extended Data Tab.~\ref{Tab:all_data}) is found to form a pre-fragment of $^{144}$Ba with a strong octupole deformation (green lines, obtained from Constrained Hartree-Fock calculations with dynamical Bardeen-Cooper-Schrieffer pairing correlations, see Methods).  {\bf b}, Confirmation of the identification of the pre-fragment by using the technique of Ref. \cite{zhang2016,sadhukhan2017} with a more general (i.e. without assuming reflection symmetry in the pre-fragment) comparison of the proton (first column) and neutron (second column)  localisation functions of the $^{258}$Fm (top half of each panel) and of the octupole constrained $^{144}$Ba (bottom half). The deformation of $^{144}$Ba used as a constraint is chosen to reproduce the nucleon localisation function close to the centre of the heavy fragment. The resulting octupole deformations of the $^{144}$Ba pre-fragment at time  T=0, 1.875, 3.75 and 4.65~zs (scission occurs at 7.3~zs) are  $\beta_3\simeq 0.14$, 0.39, 0.39 and 0.42, respectively. 
Such strong octupole deformations could not be reached in $^{132}$Sn ``doubly-magic'' nucleus without a high deformation energy cost (25~MeV for $\beta_3=0.39$), thus hindering the formation of this fragment. 
The fact that the  densities and localisation functions of deformed $^{144}$Ba match the heavy pre-fragment so well provides a clear signature of the influence of this pre-fragment prior to and at scission. }
\label{fig:prefrag} 
\end{figure*}

 \begin{figure*}[h]
\includegraphics[width=  \linewidth]{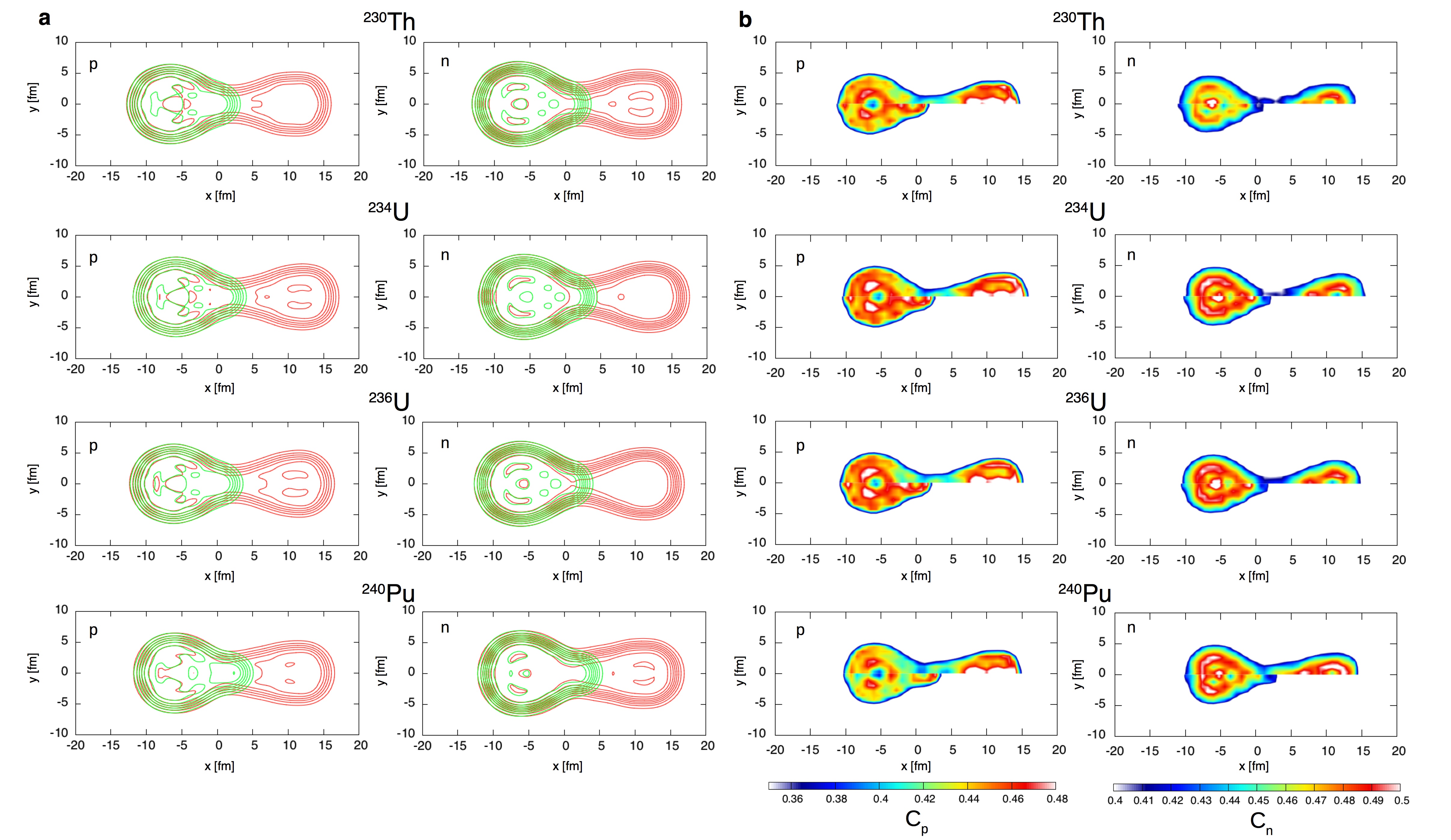}
\caption{ {\bf Identification of the heavy pre-fragment in asymmetric fission of actinides.}\\
Same as Extended Data Fig.~\ref{fig:prefrag} {\bf a} and {\bf b}, at configurations around the scission for asymmetric fission of $^{230}$Th, $^{234}$U, $^{236}$U and $^{240}$Pu.
In all four systems, the heavy fragment is identified as a $^{144}$Ba with a constrained octupole deformation corresponding respectively to $\beta_3=$0.28, 0.28, 0.27 and 0.44. The matching between deformed $^{144}$Ba densities and localisation functions with the heavy pre-fragment confirms the strong influence of octupole shell  effects associated with $Z=56$ and $N=88$ on asymmetric fission.
 }
\label{fig:prefrag_syst} 
\end{figure*}

 \begin{figure*}[t]
\includegraphics[width=  16.5cm]{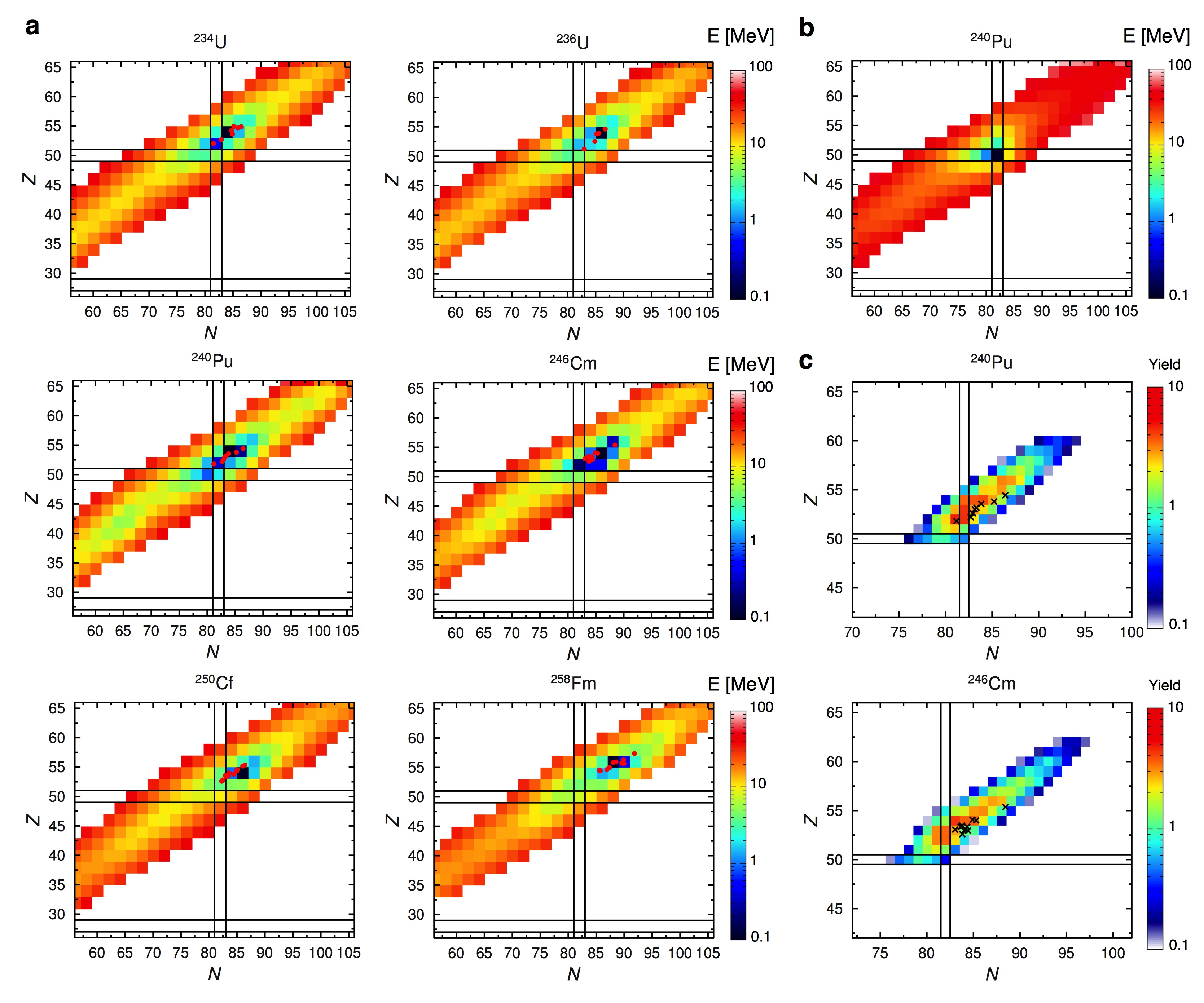}
\caption{ {\bf Effect of octupole deformation of the heavy pre-fragment on total energy at scission. }\\
{\bf a}, (logscale) To understand why the formation of a fragment in the  $^{144}$Ba region is energetically more favourable than in the $^{132}$Sn region, we calculated the total energy of the system from a simple scission-point model\cite{wilkins1976} for various mass and charge repartitions between the fragments, characterised by the number of protons $Z$ and neutrons $N$ in one fragment, and with the typical deformations of the fragments observed (in our Time-Dependent Hartree-Fock calculations with Bardeen-Cooper-Schrieffer dynamical pairing correlations, TDBCS, see Methods) at scission.
For simplicity, we only constrain the octupole deformation of the heavy fragment to be $\beta_3=0.35$ for the heavy fragment 
and  the quadrupole deformation to be $\beta_2=0.6-0.8$ for the light fragment. 
The binding energy of each deformed fragment is then computed from Constrained Hartree-Fock calculations with Bardeen-Cooper-Schrieffer pairing correlations (CHF+BCS) simulations (see Methods) and added to  the Coulomb energy between the fragments, approximated by the point like formula $e^2 Z_1Z_2 / D$ with $D=17$~fm. (As we are only interested in comparisons between different mass and charge repartitions, the strong nuclear interaction energy between the fragments is neglected as it  is not expected to vary much).
 The total energy $E(N,Z)$ is then plotted with its minimum value as reference energy for each system.  
Note that this is a simple model which does not account for finite temperature effects which could potentially damp shell effects. 
However, the damping of shell effects is expected to occur at higher excitations energies  than those involved here.
 {Despite the simplicity of this model,} the $Z$ and $N$ of the fragments obtained from the TDBCS calculations, shown by red dots, are clearly distributed around the system with minimum energy.  
{\bf b}, Same as above, but removing the constraint on the octupole deformation of the heavy fragment (only the quadrupole deformation of the light fragment is constrained). 
In this case, the formation of a $^{132}$Sn is energetically favoured. This shows that the octupole deformation of the heavy fragment induced in the fission process strongly hinders the impact of spherical shell effects at scission. 
 {\bf c}, Experimental $^{240}$Pu and $^{246}$Cm independent fission yield from Ref. \cite{BROWN20181} in logscale compared to the mean $Z$ and $N$ obtained from TDBCS calculation (black crosses). These figures show that to take into account the octupole deformation energy leads to a preference for the fragments to be formed with $Z_{\rm Heavy}\simeq54$ and overcome the effect of the spherical doubly magic $^{132}$Sn. }
 \label{fig:deform_Q} 
\end{figure*}

 \begin{figure*}[t]
\centering \includegraphics[width=  10cm]{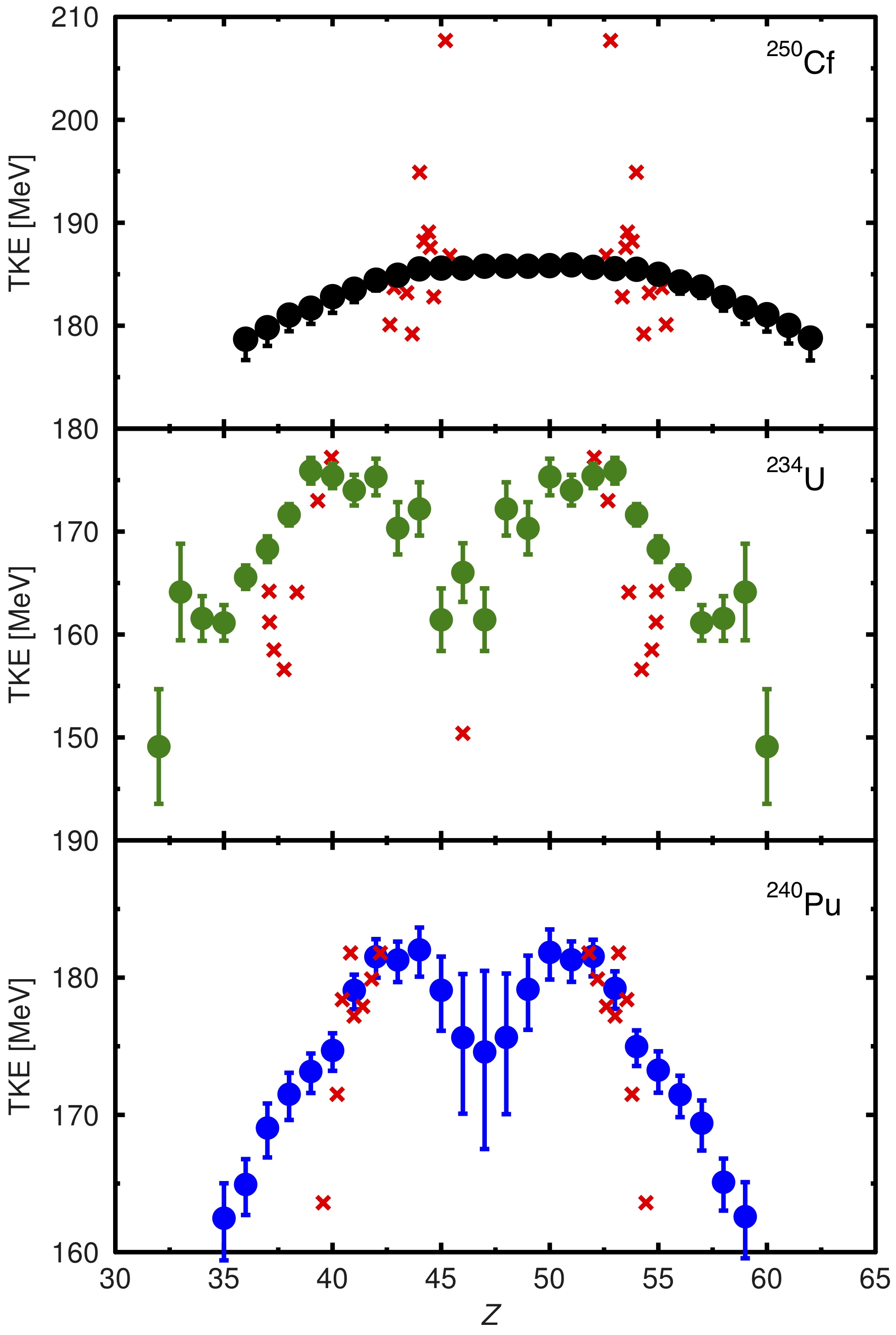}
\caption{ {\bf Total kinetic energy of the fission fragments.}\\
The total kinetic energy (TKE) obtained from  Time-Dependent Hartree-Fock calculations with Bardeen-Cooper-Schrieffer dynamical pairing correlations (TDBCS, red crosses) are compared to the average TKE from experimental data (dots) for $^{240}$Pu, $^{250}$Cf \cite{caamano2015}, 
 and $^{234}$U  \cite{bockstiegel1998}. 
 As expected from the complexity of the many-body dynamics, the results exhibit strong fluctuations 
 (typically a variation of 15-20 MeV between the lowest and highest TKE for each nucleus, i.e., of the same order of experimental fluctuations of TKE).
Nevertheless, the TKE predicted by our TDBCS calculations are essentially distributed around the experimental average TKE, 
indicating a very good agreement between  theory and  experiment. 
For consistency, we have calculated the TKE of a $^{234}$U symmetric fission mode (lowest red cross at $Z=46$ in middle panel). This calculation describes qualitatively the decrease of the TKE for symmetric fission.  
 }
\label{fig:TKE_fct_Z} 
\end{figure*}

\begin{figure*}[h]
\centering \includegraphics[width=  16.5cm]{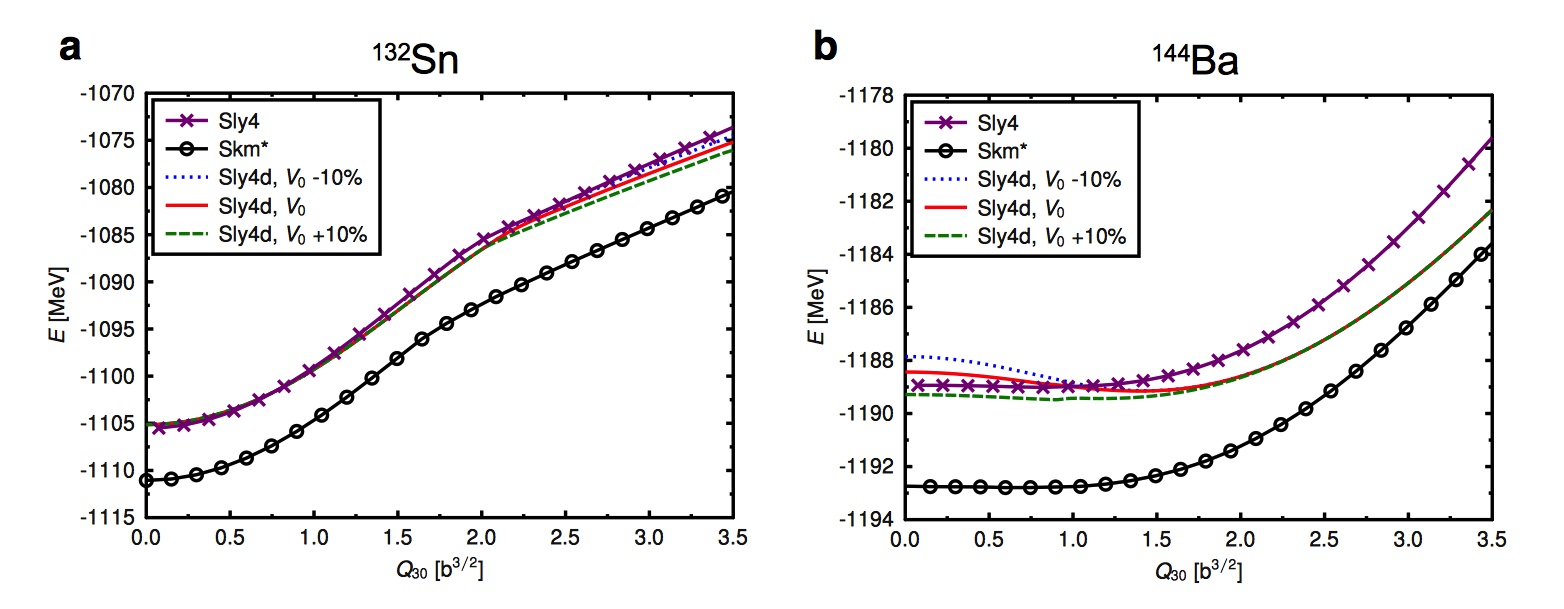}
\caption{{\bf Effect of the functional and pairing interaction }\\
Deformation energy for the $^{132}$Sn ({\bf a}) and $^{144}$Ba ({\bf b}) with different choice of functional and pairing interaction strength $V_0$ (see Methods) varying by $\pm10\%$.  The Sly4 and Skm* functional with the centre of mass correction and the Sly4d without the centre of mass correction give similar deformation energy curve. The pairing interaction can slightly change the octupole deformation of the ground state of the $^{144}$Ba. The Sly4d functional with the normal pairing interaction predicts a ground state octupole deformation $\beta_3$=0.165 that is very close to the  experimental value \cite{bucher2016} $\beta_3$=0.17(+4,-6).
 }
\label{fig:effect_interact} 
\end{figure*}


\end{document}